# Coexistence of Donor and Acceptor Hydrogen States in n-Type InN

Masaki Kobayashi[1,*], Yudai Yamashita[1], Kazuyuki Hirama[1], and Yoshitaka Taniyasu[1]

[1]*Basic Research Laboratories, NTT, Inc., 3-1 Morinosato Wakamiya, Atsugi-shi, Kanagawa 243-0198, Japan.*

**Author to whom correspondence should be addressed to**: masaki.kobayashi.brl@ntt.com

## Abstract

Hydrogen often exhibits amphoteric behavior in semiconductors, but its role is in n-type InN remains unresolved. Wurtzite InN is a narrow-gap semiconductor with high electron mobility and is therefore attractive for high-speed electronics and optoelectronic applications. Here we use hard x-ray photoemission spectroscopy (HAXPES) to probe hydrogen-related electronic structure in as-grown and post-annealed InN thin films prepared at different grown temperatures. Post annealing, which reduces the concentration of hydrogen impurities in the films, shifts the core-level spectra toward lower binding energy, consistent with a chemical-potential shift associated with the passivation of electron carriers. In the valence-band spectra, an acceptor-like in-gap feature near the valence-band maximum is suppressed after annealing. Together with the established donor-like behavior of hydrogen in InN, these results suggest that acceptor $H^-$ states coexist with donor $H^+$ states in InN. The coexistence of these opposite hydrogen charge states provides a microscopic picture of hydrogen-driven compensation in InN and highlights the amphoteric nature of hydrogen even in a highly n-type semiconductor.

## I. INTRODUCTION

Hydrogen is an unusual impurity in semiconductors. Owing to its small size, high mobility, multiple charge states ($H^+$, $H^0$, $H^-$), and strong interactions with dopants and native defects, hydrogen can either donate or compensate carriers depending on the host material and Fermi-level position [1, 2]. This amphoteric behavior underlies a wide range of phenomena, including passivation of Mg acceptors in GaN:Mg [3, 4, 5, 6], dopant deactivation in Si [7, 8, 9, 10], and changes in the electrical properties of conductive transparent oxides [11, 12, 13, 14, 15, 16].

Because hydrogen can adopt different charge states in semiconductors, it is therefore a central factor governing the electronic properties of semiconductors.

InN provides a particularly intriguing case. Wurtzite InN is a narrow bandgap semiconductor with a bandgap energy of ~0.65 eV [17, 18, 19] and theoretically predicted electron mobilities as high as 14,000 $cm^2/Vs$ have been reported [20]. These properties make it attractive for high-speed electronics. A major obstacle, however, is that InN typically exhibits degenerate n-type conductivity with a large residual electron concentration ($n$), which hinders practical device applications. Possible origins of the residual $n$ include point defects [N vacancy ($V_N$) and dopant complexes] [21, 22, 23], extrinsic impurities (H, O, etc) [24, 25, 26, 27], and threading dislocations [28, 29]. Previous theoretical calculations indicate the hydrogen in InN acts as a donor and can occupy several stable locations in InN [27], and experiments have identified shallow donor states of hydrogen [25, 26]. More generally, however, hydrogen tends to act as a donor ($H^+$) in p-type materials and as an acceptor in n-type materials [5]. The donor-like behavior reported for hydrogen in n-type InN therefore appears to be at least partly at odds with this general tendency, making a direct probe of the hydrogen-related electronic states particularly important.

To clarify the electronic states of hydrogen in InN, we performed hard x-ray photoemission spectroscopy (HAXPES) on as-grown (AG) and post-annealed (PA) InN thin films prepared at different growth temperatures ($T_s$). HAXPES is well suited to this problem because its probing depth ($l$), on the order of several tenth nanometer, is much larger than that of conventional photoemission spectroscopy (PES) with laboratory X-ray sources ($l \sim 5$ nm), making it substantially less sensitive to surface contributions [30, 31, 32, 33]. The core-level spectra show that the post-annealing shifts the chemical potential in a direction consistent with compensating the residual electrons, even though the concentration of donor-like $V_N$ is slightly increased by annealing. The valence-band (VB) spectra further reveal an acceptor-like impurity feature near the valence-band maximum (VBM) that is suppressed after annealing, suggesting a hydrogen-related origin. Together, these observations indicate that the donor-like and acceptor-like hydrogen states coexist in n-type InN thin films. In the following, we discuss the implications of this coexistence for the role of hydrogen in InN.

## II. METHODS

Wurtzite InN thin layers were grown on $GaN/Al_2O_3(0001)$ substrates by Metalorganic

Vapor Phase Epitaxy (MOVPE) with different substrate temperatures from 450 to 565 °C. Trimethylindium and $NH_3$ were used as precursors, and $H_2$ and $N_2$ served as carrier gases. The total pressure during the growth was 1500 hPa. All samples were treated with HCl for 1 min to remove surface In droplets. The crystal quality of the InN layers was evaluated by x-ray diffraction (XRD) and cross-sectional transmission electron microscopy. Details of the growth procedure are given elsewhere [34]. The films were post-annealed at 450 °C for 40 min in nitrogen at atmospheric pressure to reduce the hydrogen concentration. The electron carrier concentration $n$ of the AG and PA films was determined by Hall-effect measurements.

HAXPES measurements were performed at beamline BL09U of NanoTerasu. The incident photon energy ($h\nu$) was 6 keV, corresponding to an inelastic mean free path $\lambda$ was estimated to be ~7.0 nm [35], where the practical probing depth $l$ is approximately three times larger than $\lambda$; $l$ ~ 21 nm. Spectra were recorded with an R4000 hemispherical photoelectron analyzer (Scienta-Omicron Co.Ltd.) under an ultra-high vacuum below $1 \times 10^{-7}$ Pa at room temperature. The monochromator resolution $E/\Delta E$ was better than 20,000, and the total energy resolution was about 270 meV. The beam spot size at the surface of the sample was $12.5 \times 3$ $\mu m^2$. The binding energy ($E_B$) was calibrated by measuring the Fermi level ($E_F$) of a gold film electrically contacted to the sample.

## III. RESULTD AND DISCUSSION

We first summarize the effect of post-annealing on the electrical properties of the InN thin films. Figure 1(a) shows electron concentration $n$ as a function of $T_s$ for the AG and PA InN thin films. Post annealing reduces $n$, and the reduction becomes larger with increasing $T_s$. Figure 1(b) shows depth profiles of H and O concentrations in the film grown at $T_s$ = 565 °C, as measured by secondary-ion mass spectroscopy (SIMS) using $Cs^+$ ions. In this work, concentrations were evaluated using GaN standard. We estimate uncertainty due to material difference between the standard and the target sample to be a factor of 2–3 as reported for the estimation of H and O concentrations in ScN with GaN standard [36]. Annealing reduces the H concentration from approximately $10^{21}$ to $10^{20}$ $cm^{-3}$, whereas the O concentration remains nearly unchanged. Similar trends were inferred for other films. It should be mentioned here that SIMS signals indicating high concentrations in the first ~20 nm, which corresponds to $l$ for $h\nu$ = 6 keV, are likely inaccurate due to effects of adsorbed molecules, oxide layers, and surface roughness. The approximately one-order-of-magnitude reduction in H concentration discussed here is evaluated from the steady-state plateau region. Because annealing primarily reduces the H concentration and $n$ in InN, these results suggest that donor-like H is reduced by annealing. It is also noteworthy that most of incorporated H is electrically inactive in InN, because the H concentration (~$10^{21}$ cm$^{-3}$) is much larger than $n$ (~$10^{19}$ cm$^{-3}$), implying that only a few percent of the incorporated H contribute directly to the conductivity.

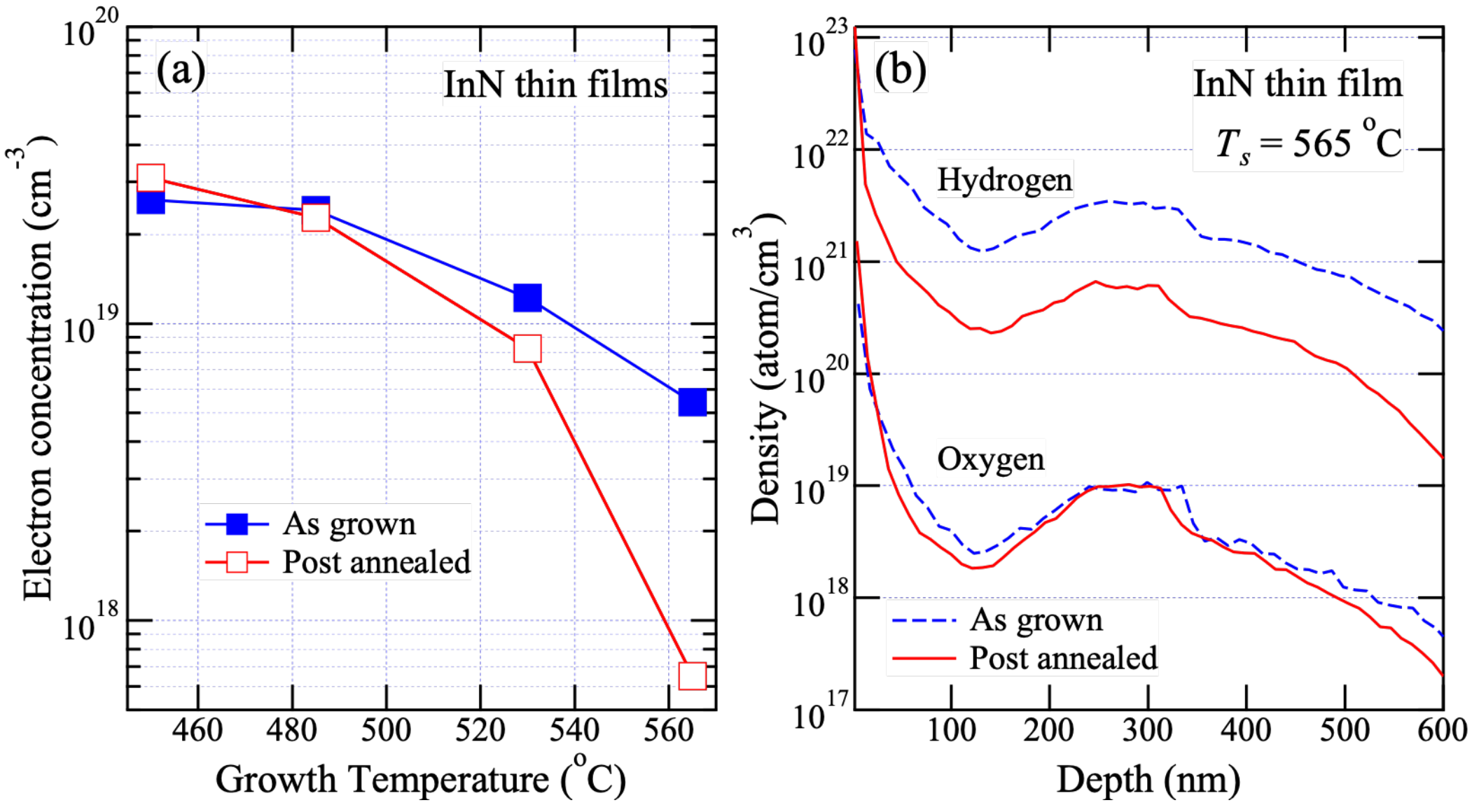

FIG. 1. Transport property of InN thin films grown at different growth temperatures. (a) Electron concentration as a function of growth temperature, determined by Hall measurements. (b) Depth profiles of H and O impurities in the film grown at 565 °C, measured by SIMS. Whereas the O concentration remains nearly unchanged, the H concentration is reduced by approximately one order of magnitude after post-annealing. In both panels, the vertical axis is logarithmic.

## A. In and N core-level spectra

Figure 2(a) shows the wide-range spectra of the PA InN thin films, including the In 3$d$ and N 1$s$ core-level peaks, normalized by the In 3$d_{5/2}$ peak height. The relative N 1$s$ peak intensity decreases with increasing $T_s$, as shown in Fig. 2(b), suggesting that the N/In ratio becomes smaller for films grown at higher temperatures. Figure 2(c) plots the N 1$s$ intensity relative to In 3$d_{5/2}$ as a function of $T_s$ for both AG and PA films. The decrease in N content with increasing $T_s$ is qualitatively similar before and after annealing. Whereas the relative N 1$s$ intensity is nearly unchanged by annealing for films grown below 485 °C, it decreases after annealing for films grown above 530 °C, even under the same annealing conditions. These trends suggest that $V_N$ are more readily created at high $T_s$ and are further increased by annealing. If the N 1*s* intensity is taken as proportional to the N content, the corresponding change in $V_N$ is less than 10%. Combined with the decrease in *n* after annealing for the film at $T_s$ = 565 °C, this result suggests that the electrons donated by the increase in $V_N$ are fully passivated and that the native electron carriers are further partially compensated in the n-type InN film.

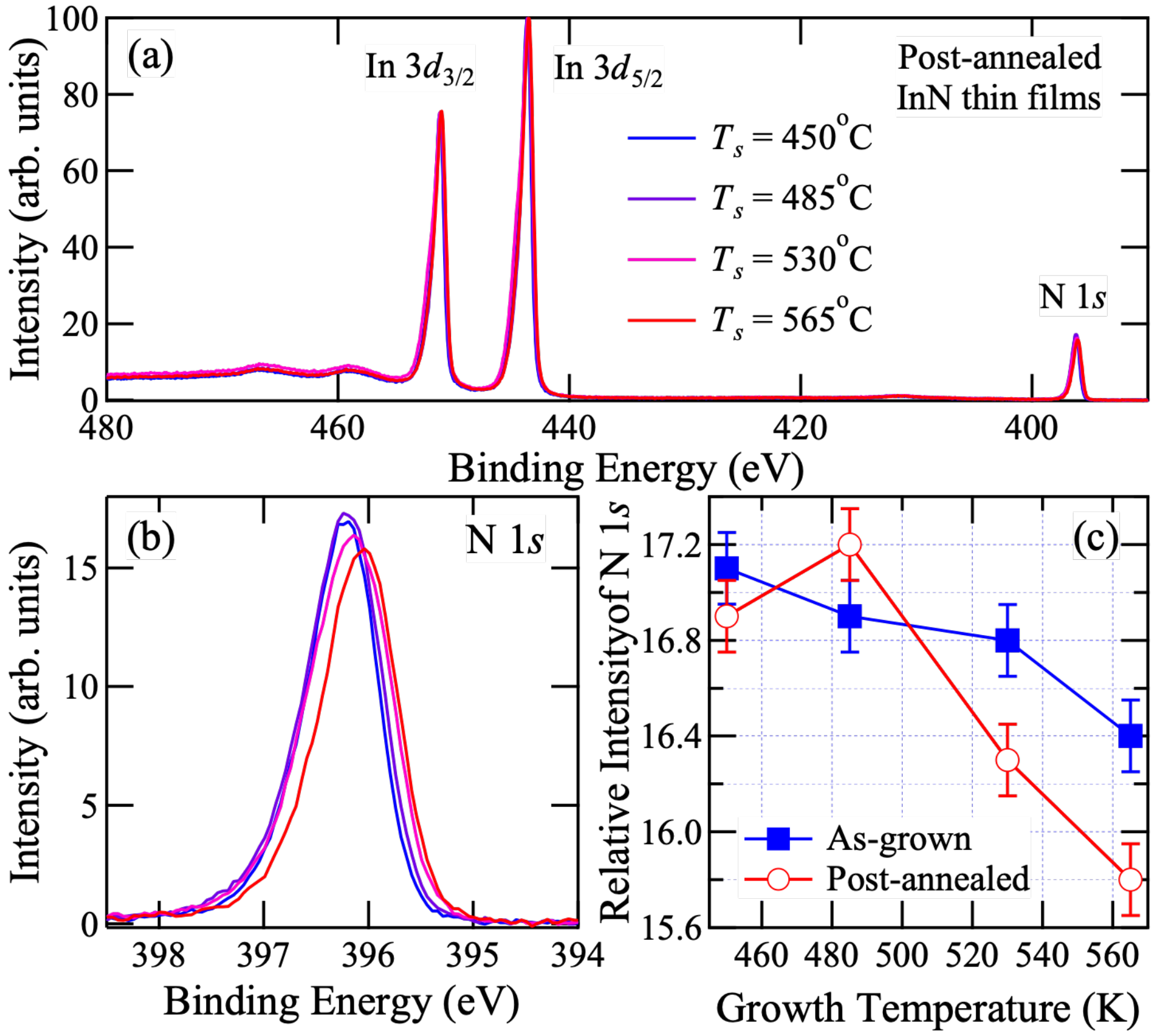

FIG. 2. In 3*d* and N 1*s* core-level spectra of InN thin films grown at different growth temperatures. The spectra are normalized to the In 3$d_{5/2}$ peak height. (a) Wide-range spectra of the post-annealed thin films. Here, the spectra are normalized so that the maximum intensity is 100. (b) Enlarged view around the N 1*s* core level on the same intensity scale as in (a). (c) Relative N 1*s* intensity normalized to the In 3$d_{5/2}$ peak, plotted as a function of growth temperature.

Figure 3 shows the In 3$d_{5/2}$ spectra of the AG and PA InN thin films, together with difference spectra obtained by subtracting the AG spectra from the PA spectra. For all films except for $T_s$ = 450 °C, the main peak shifts to lower $E_B$ after annealing, consistent with a chemical-potential shift toward lower electron concentration. In addition to this common shift, subtle line-shape changes are observed on the higher $E_B$ side, especially for the $T_s$ = 530 °C sample, where a clear shoulder structure appears after annealing. This shoulder structure likely contains contributions from extrinsic $InO_x$ (*x* is larger than 3/2) near the surface. Indeed, a weak peak from an In secondary phase was observed by XRD only for the $T_s$ = 530 °C sample. Annealing may therefore oxidize a surface In component and/or enhances the In secondary phase. Because of the long probing depth of HAXPES, however, the main peak is still expected to reflect the bulk electronic state of the InN film, and the principal chemical shift is likely intrinsic (see also Appendix I for XPS data taken by Mg-$K\alpha$ x-ray source with shallower probing depth compared to that of HAXPES).

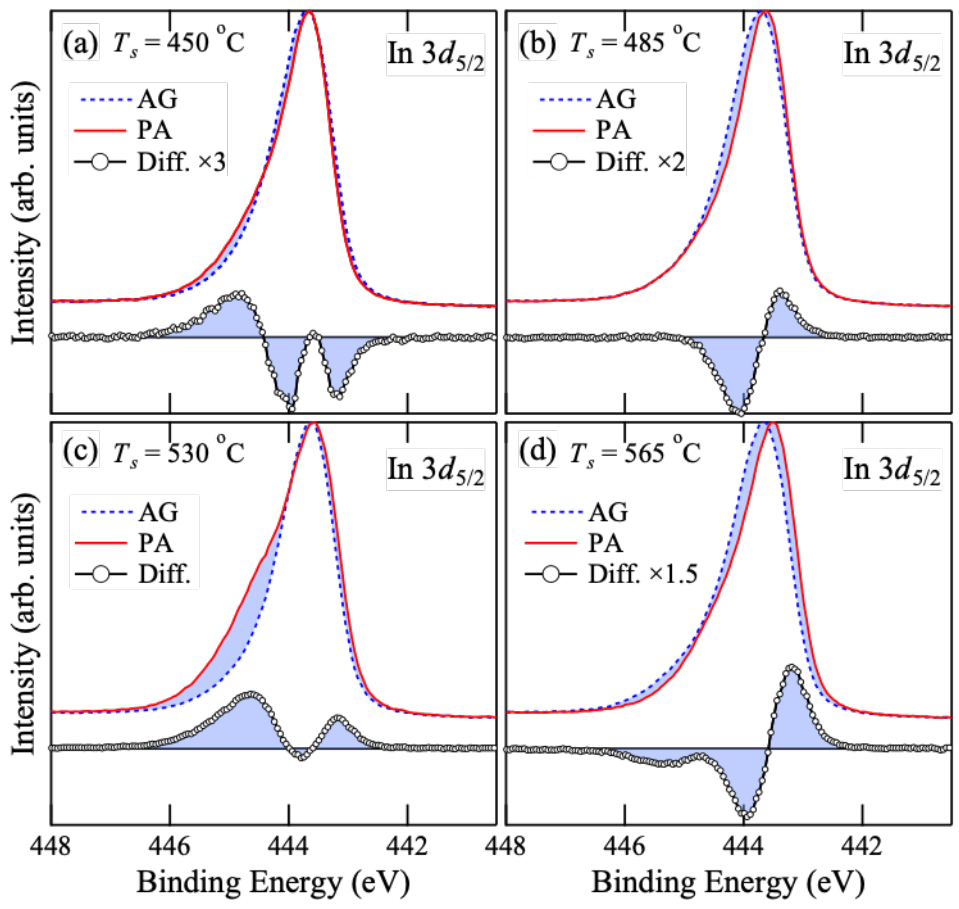

FIG. 3. In $3d_{5/2}$ spectra of the as-grown (AS) and post-annealed (PA) InN thin films. The spectra are normalized to the height of the main peak. (a)–(d) Spectra of films grown at $T_s$ = 450, 485, 530, and 565 °C, respectively. The shaded areas indicate the spectral differences, and the difference spectra between the AG and PA films are also shown.

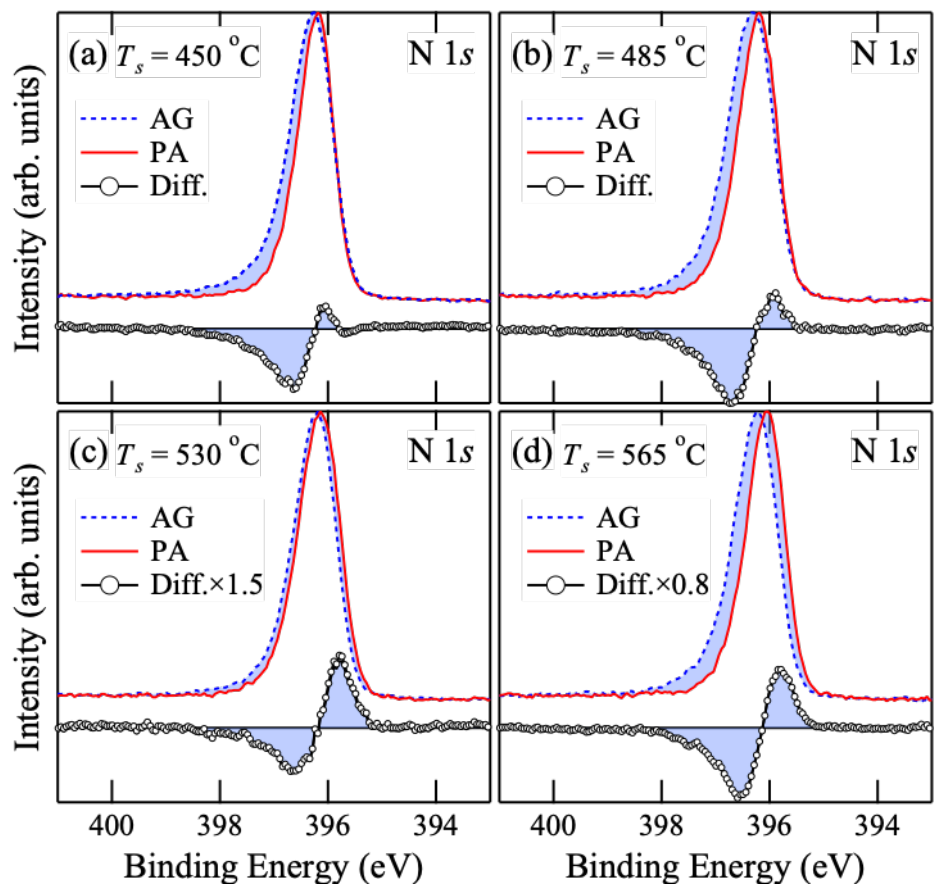

FIG. 4. N 1$s$ spectra of the AS and PA InN thin films. The spectra are normalized to the height of the main peak. (a)–(d) Spectra of films grown at $T_s$ = 450, 485, 530, and 565 °C, respectively. The shaded areas indicate the spectral differences, and the difference spectra between the AG and PA films are also shown.

Figure 4 compares the N 1$s$ spectra before and after annealing. All spectra exhibit asymmetric line shapes with tail structures on the higher-$E_B$ side of the main peak. After annealing, the main peak shifts toward lower $E_B$, similar to the In 3$d$ spectra, and the tail structure decreases in intensity for all $T_s$. Notably, the reduction of the tail structures in the N 1$s$ spectra is opposite to the enhancement of the higher $E_B$ components in the In 3$d$ spectra, implying that annealing reduces an N-related component with a higher effective valence than $N^{3-}$ in InN.

**B. Spectral shape analysis**

Figures 5(a) and 5(b) summarize the main peak positions of the In $3p_{3/2}$ and N 1$s$ spectra, respectively. The shifts of the In $3p_{3/2}$ and the N 1$s$ are nearly identical, indicating a common chemical-potential shift rather than an element-specific chemical shift. While the peak positions of the AG films vary only weakly with $T_s$, those of the PA films shift monotonically to lower $E_B$ with increasing $T_s$. This $T_s$ dependence is consistent with the trend of $n$ in Fig. 1(a). These results indicate that the spectral shifts reflect the chemical-potential shift associated with the carrier concentration of the InN thin films. Both the core levels in the PA films lie at lower $E_B$ than in the corresponding AG films, consistent with a chemical-potential shift toward the VBM and reduced $n$ after annealing. This result is opposite to the increase in $V_N$, which acts as triple donor and is inferred to grow at higher $T_s$ [see Fig. 2(c)], implying that the H species reduced by annealing predominantly contributes to electron doping.

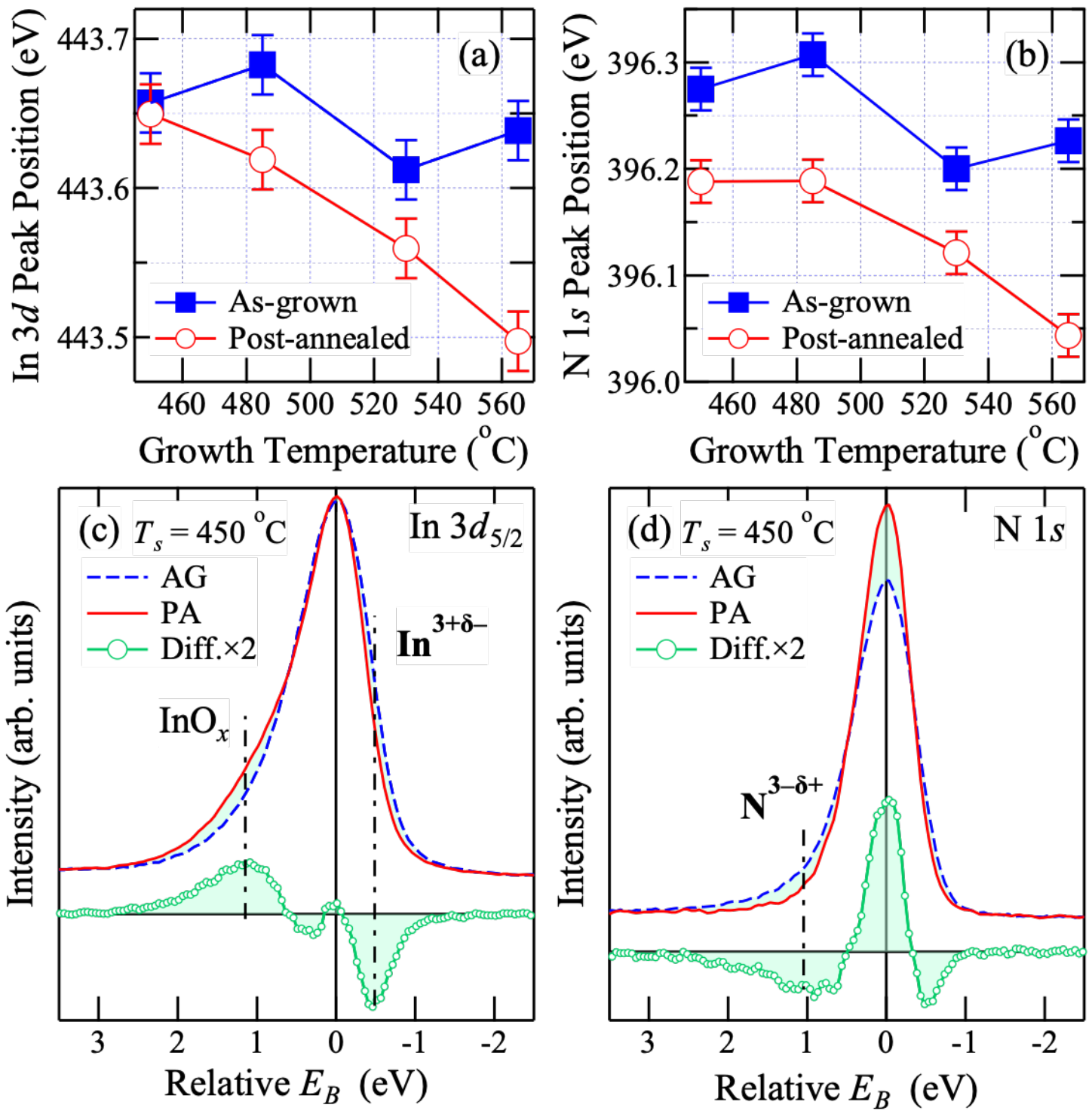

FIG. 5. Spectral-shape analysis for the core-level spectra of the InN thin films. (a),(b) Main-peak positions of In 3$d_{5/2}$ and N 1$s$, respectively, plotted as a function of growth temperature. (c),(d) Comparison of the In 3$d_{5/2}$ and N 1$s$ spectra, respectively, between the AG and PA films grown at 450 °C. Here, the spectra are shifted to align the main-peak positions and are normalized by area. In (c) and (d), the shaded areas denote the spectral differences. The vertical dash-dotted lines are guides for eyes.

To identify the line-shape changes more clearly, we performed a spectral-shape analysis. The AG and PA spectra were aligned to the main-peak position to remove the chemical-potential shifts shown in Figs. 5(a) and 5(b). Figures 5(c) and 5(d) show the representative analyses of the In 3$d_{5/2}$ and N 1$s$ spectra for the film grown at $T_s$ = 450 °C, respectively, plotted as a function of relative $E_B$ referenced to the peak position, with spectra normalized by area (all datasets are shown in Appendix II and III). In Fig. 5(c), the difference spectrum of the In $3d_{5/2}$ exhibits a negative feature below the main peak and a positive feature above it. The negative (positive) feature in the difference spectrum indicates a component that is suppressed (enhanced) by annealing. Because the main peak is assigned to the $In^{3+}$–$N^{3-}$ bonding state (formal charges within ionic picture; the effective valence needs not to be integer because of covalency), the component on the lower $E_B$ side indicates an In species of lower effective valence; 3+δ–. Since this component is reduced by annealing, it is plausibly associated with In–H bonding, i.e., a minority In–H-related state in the AG films. The In 3*p* spectra therefore appear to consist of the main In–N, an extrinsic $InO_x$ ($x >$ 3/2), and a weaker In–H-related components. The presence of the In–H-related component with the lower effective valence suggests H impurities that trap electron carriers, corresponding to an acceptor-like $H^-$ state located around the $In^{3+}$ sites. Although the acceptor $H^-$ state has not been theoretically predicted for InN, such a state is qualitatively consistent with the general trend of impurities in semiconductors to compensate the active carriers toward charge neutrality.

In the N 1*s* difference spectra, a negative component appears above the main peak, although a feature at zero energy remains because of the normalization. This component is plausibly associated with the N–H bonding because it is reduced by annealing and has a higher effective valence than the main In–N bonding state; 3–δ+. This result therefore suggests donor-like $H^+$ species located around the $N^{3-}$ sites. Donor-like H configurations proposed for InN include H at $V_N$, interstitial H, and bond-centered H [27], and H-related donor complexes such as the solitary-cation model [37]. However, the present spectra do not uniquely determine the microscopic H site, and we therefore regard this configuration as a plausible model rather than a definitive assignment. Shallow H-related donor states have also been reported experimentally [25, 26]. The analysis indicates the coexistence of the donor-like $H^+$ and acceptor-like $H^-$ states in n-type InN. It should be mentioned here that it seems to be difficult to observe these H states separately within the experimental accuracy because of the energy resolution of ~270 meV together with intrinsic lifetime broadening. Thermochemical data for gas-phase species relevant to III-nitride growth show that $NH_3$ is thermodynamically stable under standard conditions (Gibbs free energy of formation: –16.4 kJ/mol) [38], whereas monomeric $InH_3$ is much less stable and is discussed as a labile species [39, 40]. This trend is qualitatively consistent with donor-like H associated with N-related environment being more thermally robust than acceptor-like H associated with In-related environment. These considerations suggest that donor $H^+$ impurities are likely to dominate the conductivity and more thermally stable than acceptor $H^-$ state in InN.

Based on the findings, let’s discuss possible defects in InN, other than hydrogen,

including $V_N$, indium vacancy ($V_{In}$), interstitial indium ($In_{int}$), and excess oxygen ($O_{ex}$). Here, $V_N$, $In_{int}$, and $O_{ex}$ act as donor, whereas $V_{In}$ acts as acceptor. Because of n-type nature of InN, majority of native defects are expected to be donor, suggesting the amount of $V_{In}$ is thought to be minimal. From comparison between the HAXPES and XPS spectra (see Supplemental Material), excess In ions are mainly located near the surface, not in the bulk. The intensity of O 1*s* spectra suggest that the concentration of oxygen near the surface increases upon annealing (not shown). Therefore, although both donor defects $V_N$ and $O_{ex}$ increase upon annealing, the observations of the decrease of the electron concentration by annealing indicate that the hydrogen impurity dominantly affects the electronic properties of the InN thin films rather than the possible defects.

**C. Valence-band spectra**

Figure 6(a) shows the VB spectra of the PA InN films grown at different $T_s$. Although the detailed line shapes vary slightly, the overall spectra are similar among the samples. At first glance, the VBM appears near $E_B \sim 1$ eV. Figure 6(b) enlarges the near-$E_F$ region and shows weak but finite spectral weight crossing $E_F$. Given n-type nature of the InN films, this small spectral weight is attributed to the conduction band of InN. Note that the leading edge around $E_B \sim 1.0$ eV gradually shifts to lower $E_B$ with increasing $T_s$, consistent with the $T_s$ dependence of the electron concentration [Fig. 1(a)]. In addition, the valleys between the valence and conduction bands becomes deeper with decreasing $T_s$. Figures 6(c)–(f) compare the near-$E_F$ spectra of the AG and PA films. Annealing mainly affects the depths of this valley, i.e., the in-gap spectral weight, for the films grown at $T_s$ = 450 °C and 485 °C, whereas much smaller changes are observed for the films grown at $T_s$ = 530 °C and 565 °C. The in-gap spectral weight was obtained by subtracting the PA spectrum from the AG spectrum after accounting for the chemical-potential shift, and the VBM was estimated by linear extrapolation of the VB leading edge, as shown in Fig. 6(g). Here, the spectra are normalized so that the VB slops below $E_B \sim 1.5$ eV coincide. For the film grown at $T_s$ = 450 °C, the in-gap spectral weight lies near the VBM and extends into the gap region, but it has no intensity at $E_F$ and therefore does not contribute directly to conductivity. Figure 6(f) summarize the in-gap spectral weight for the films with $T_s$ = 450, 485, and 530 °C. Its intensity decreases with increasing $T_s$, while the negative spectral weight likely reflects the reduced occupation of the conduction band after annealing. Because these in-gap states are suppressed by annealing, they are likely related to the hydrogen-derived acceptor-like states. In contrast to donor $H^+$ ($1s^0$), whose empty 1*s* state is not directly visible in photoemission, acceptor $H^-$ ($1s^2$) is occupied and can contribute to the observed VB spectral weight. These observations therefore provide additional evidence for the existence of the acceptor $H^-$ states in the InN thin films.

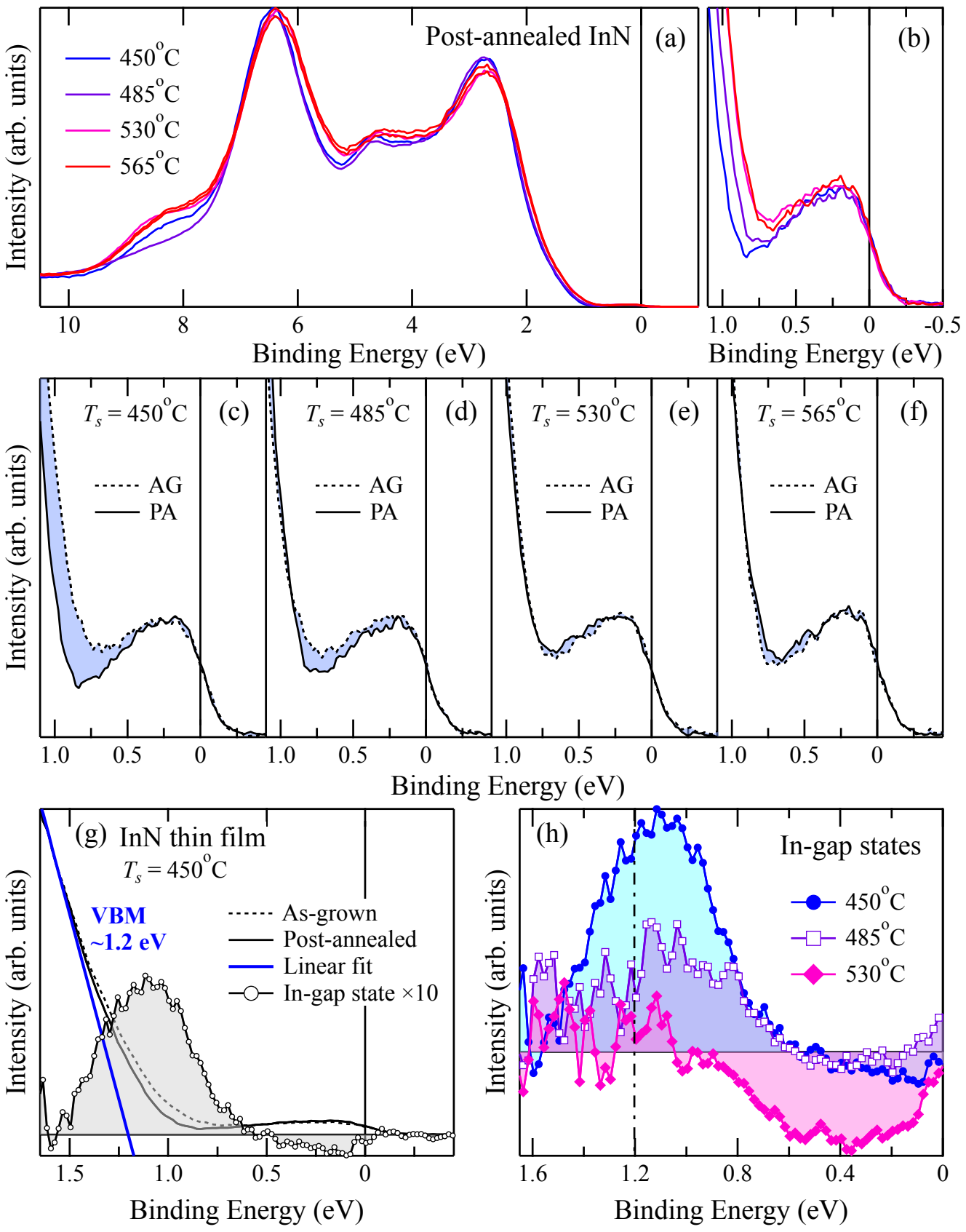

FIG. 6. Valence-band (VB) spectra of InN thin films. (a) VB spectra of PA InN films grown at different $T_s$, normalized by area. (b) Enlarged view near $E_F$, normalized to the peak height. (c)–(f) Near-$E_F$ spectra of the films grown at $T_s$ = 450, 485, 530, and 565 °C, respectively. (g) Estimations of the valence-band maximum (VBM) and the in-gap spectral weight for the InN thin film grown at $T_s$ = 450 °C. (h) In-gap spectral weight for the films with $T_s$ = 450, 485, and 530 °C. The vertical dash-dotted line indicates VBM. In (g) and (h), the horizontal sold lines denote zero intensity.

## IV. CONCLUSION

In conclusion, we used HAXPES to investigate hydrogen-related electronic states in InN thin films grown at different growth temperature before and after post-annealing. Annealing reduces the H concentration by about one order of magnitude ($10^{21}$ to $10^{20}$ $cm^{-3}$) and lowers the electron concentration $n$ from the $10^{19}$ $cm^{-3}$ to $10^{18}$ $cm^{-3}$ range. The In 3$p$ and N 1$s$ peaks in the PA films shift systematically with $T_s$, in contrast to less shifts in the AG films, consistent with the $T_s$ dependence of $n$. The shifts reflect the chemical-potential shift associated with reduced electron concentration. Line-shape analysis further identifies an In–H-related component on the lower $E_B$ side of the In 3*p* spectra, and an N–H-related component on the higher $E_B$ side of the N 1*s* spectra. Together with the disappearance of an in-gap feature near the VBM after annealing, these results suggest that acceptor-like $H^-$ states, not expected to exist hitherto, coexist with donor-like $H^+$ states in n-type InN. The existence of the acceptor impurities in an n-type semiconductor is plausible in light of the general tendency for charged impurities to compensate active carriers toward charge neutrality. The coexistence provides a more complete picture of hydrogen in InN than a donor-only description and points to hydrogen-driven compensation as an important part of the residual-carrier problem. Because donor $H^+$ appears to be more thermally stable than acceptor $H^-$ state, donor-like $H^+$ state is still likely to dominate the conductivity, consistent with previous studies of H impurity in InN. More broadly, the present HAXPES approach should be useful for disentangling defects and impurity-related states in semiconductors.

**ACKNOWLEDGMENTS**

S. Kobari and Y. Watanabe are acknowledged for the technical support of the HAXPES experiment at NanoTerasu BL09U.

## Appendix

### I. XPS measurements with Mg-*Kα* x-ray source

To access the surface states, experimentally we have measured the InN thin films with a Mg-*Kα* x-ray source ($h\nu$ = 1253.6 eV), for which the probing depth is approximately 5 nm. Figures S1(a) and S2(b) show the In 3*d* and N 1*s* core-level spectra of n-type InN thin films, respectively. It should be mentioned here that the surface conditions of the films were likely getting worse due to elapsed time from the HAXPES measurements and the pieces of the films were different from these measured by HAXPES. Then, additional extrinsic effects were included in these spectra. As shown in the In 3*d* spectra (Fig. S1(a)), the oxidized In components seem to be dominant in the annealed films, while the oxidized and InN components can be seen in the spectra taken in the as-grown films. The observations suggest that annealing leads to thick oxidized In layer near the surface, consistent with the oxidized In component observed in the HAXPES measurements. As for the N 1*s* spectra shown in Fig. S1(b), the higher binding energy ($E_B$) components are reduced by annealing in the films with $T_a$ = 450 °C and 485 °C. On the other hand, a prominent peak appears in higher $E_B$ side in the films with $T_a$ = 530 °C and 565 °C. A possible interpretation is that nitrogen ions catch hydrogen impurities diffused toward the surface. These surface-enhanced features are weak or absent in the HAXPES spectra, demonstrating that HAXPES substantially suppresses, but does not completely eliminate, surface contributions. This is consistent with our statement that the HAXPES spectra mainly reflect the intrinsic bulk electronic states of the InN thin films.

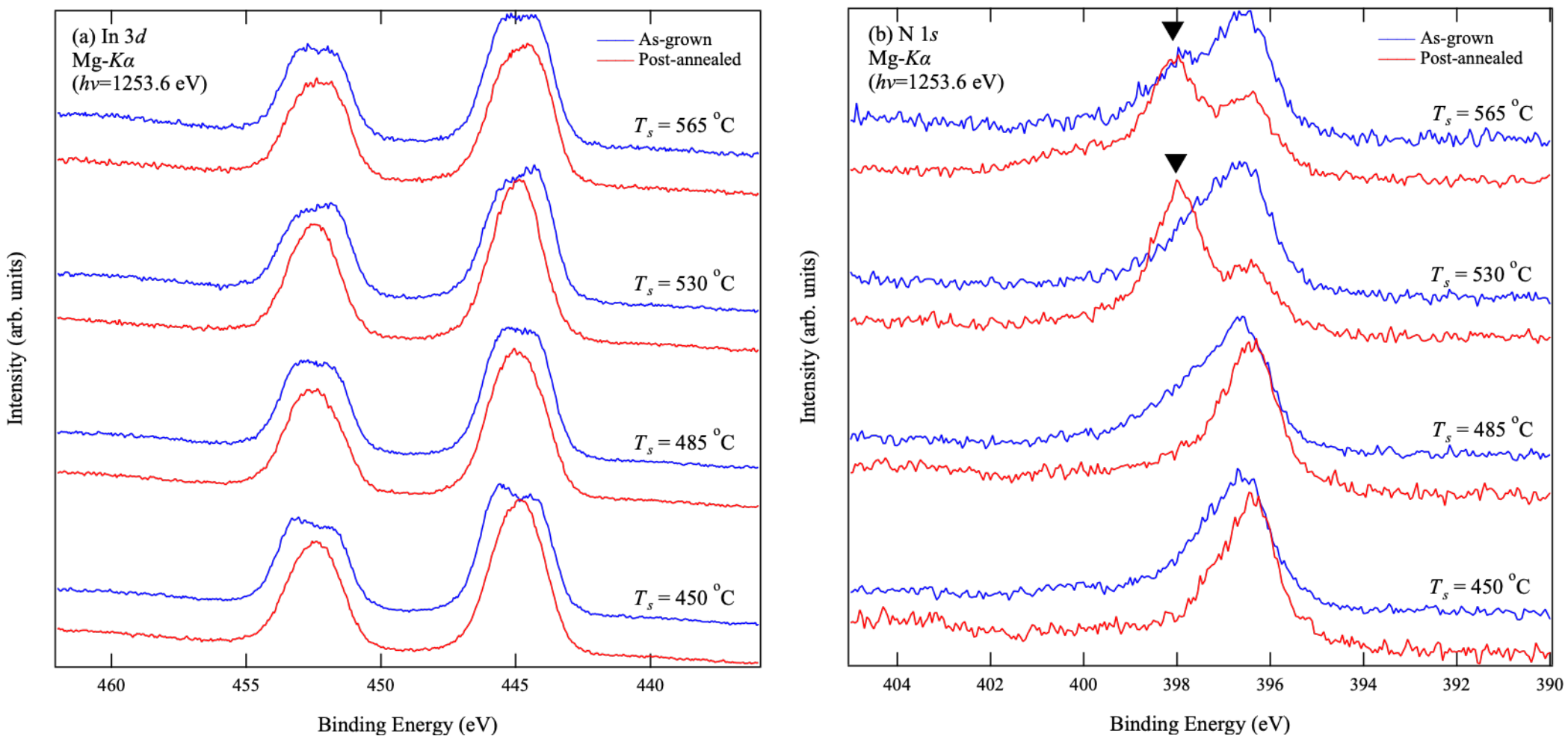

FIG. S1. Core-level spectra of InN thin films taken by Mg-*Kα* x-ray source ($h\nu$ = 1253.6 eV). (a) In 3*d* spectra. (b) N 1*s* spectra. Down-triangles denote extra peaks appeared after annealing.

## II. Comparison of the line shapes of the In 3*d* spectra

Figure S2 shows the spectral-shape analysis of the In $3d_{5/2}$ spectra of the InN thin films. As discussed in the main text, the component below the main peak is consistent with an In–H-related contribution, whereas the higher-binding-energy side likely includes an extrinsic $InO_x$ ($x > 3/2$) contribution. Although the intensity of the extrinsic component depends on the growth condition, the In–H-related component monotonically decreases with increasing the growth temperature ($T_s$). It should also be noted that the difference spectra are nearly identical except for the film grown at $T_s$ = 530 °C, for which the oxidized In contribution is relatively large. Because the In–H-related state is expected to be thermally unstable, both growth at higher $T_s$ and post-annealing likely reduce the amount of the acceptor-like $H^-$ impurities.

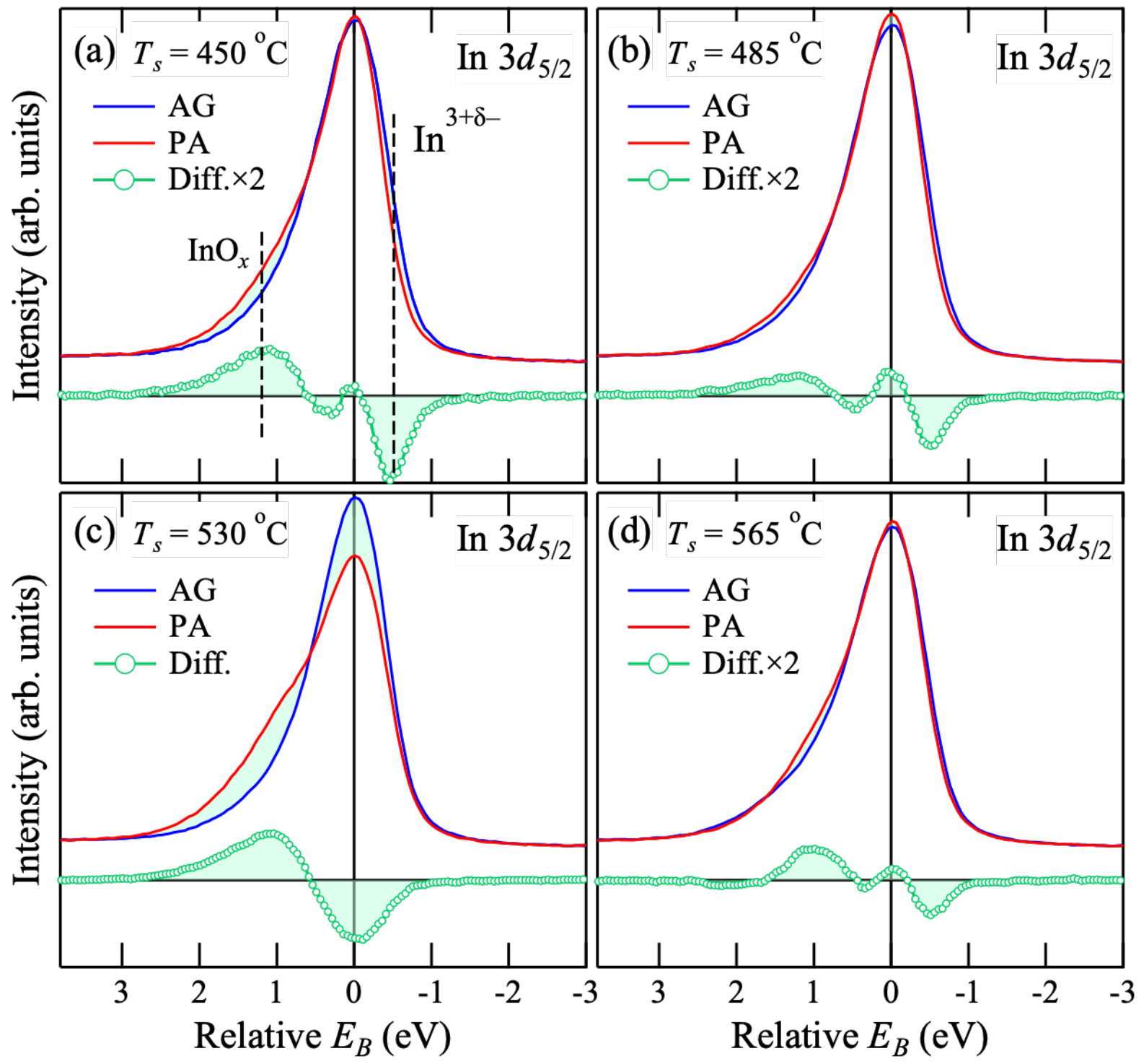

FIG. S2. Spectral-shape analysis for the In $3d_{5/2}$ core-level spectra of the InN thin films. (a)–(d) Comparison of the In $3d_{5/2}$ core-level spectra between the as-grown (AG) and post-annealed (PA) film grown at 450, 485, 530, and 565 °C, respectively. Here, the spectra are shifted to align the main-peak positions and are normalized to the spectral areas. The shaded areas denote the spectral differences. The vertical dashed lines are guides for eyes.

## III. Comparison of the spectral shapes of the N 1*s* spectra

Figure S3 shows the spectral-shape analysis of the N 1*s* spectra of the InN thin films. The difference spectra are nearly identical among the samples. In contrast to the In–H-related contribution, the area of the tail structure above the main peak is nearly unchanged with $T_s$. This behavior is consistent with the greater thermal stability of N–H bonding states, which are expected to contribute to the donor-like hydrogen component. The relative intensities of the In–H and the N–H-related features nevertheless differ among the films. If acceptor-like $H^-$ compensates electrons in InN, the present trend suggests that annealing removes comparable amounts of acceptor-like and donor-like H impurities in the film grown at $T_s$ = 450 °C, whereas a large fraction of donor-like H impurity is removed in the film grown at $T_s$ = 565 °C. This picture is qualitatively consistent with the $T_s$ dependence of the electron carrier concentration $n$ [Fig. 1(a)].

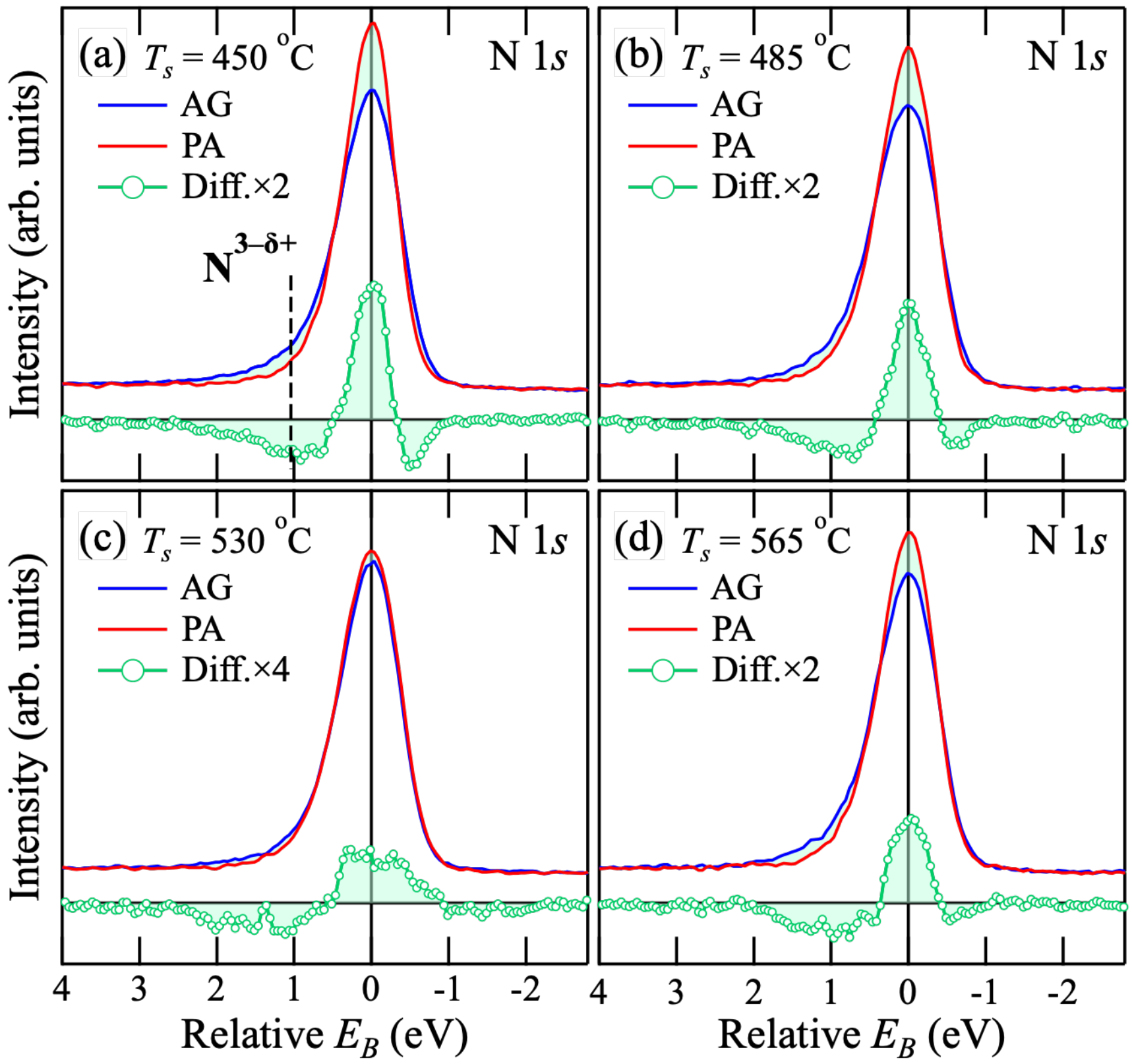

FIG. S3. Spectral-shape analysis for the N 1$s$ core-level spectra of the InN thin films. (a)–(d) Comparison of the N 1$s$ core-level spectra between the AG and PA film grown at 450, 485, 530, and 565 °C, respectively. Here, the spectra are shifted to align the main-peak positions and are normalized to the spectral areas. The shaded areas denote the spectral differences. The vertical dashed lines are guides for eyes.